\documentclass[preprint,eqsecnum,nofootinbib,superscriptaddress,prd,showpacs]{revtex4}
\def\be{\nopagebreak[3]\begin{equation}}
        \def\ee{\end{equation}}
        \def\ba{\nopagebreak[3]\begin{eqnarray}}
        \def\ea{\end{eqnarray}}
\begin{document}
\title{Towards the deformation quantization of linearized gravity }
\author{Hernando Quevedo}\email{quevedo@physics.ucdavis.edu}
\affiliation{Instituto de Ciencias Nucleares\\
Universidad Nacional Aut\'onoma de M\'exico\\
A. Postal 70-543, M\'exico D.F. 04510, MEXICO \\
}
\affiliation{Department of Physics\\
University of California\\
Davis, CA 95616, USA\\
}
\author{Julio G. Tafoya}\email{antonio@nuclecu.unam.mx}
\affiliation{Instituto de Ciencias Nucleares\\
Universidad Nacional Aut\'onoma de M\'exico\\
A. Postal 70-543, M\'exico D.F. 04510, MEXICO \\
}

\begin{abstract}

We present a
 first attempt to apply the approach of deformation 
quantization to linearized Einstein's equations.
We use the analogy with Maxwell equations to 
derive the field equations of linearized gravity
from a modified Maxwell Lagrangian which allows
the construction of a Hamiltonian in the standard
way. The deformation quantization procedure 
for free fields is applied to this Hamiltonian.
As a result we obtain the complete set of 
quantum states and its discrete spectrum.
\end{abstract}
 \maketitle

\section{Introduction}
\label{sec1}

One of the main differences between the states 
of a classical system and those of a quantum system 
consists in that the latter cannot be represented
as points in phase space. In the canonical quantization
approach this fact is taken into account by considering
states as eigenfunctions of operators which act on
Hilbert space. The eigenvalues of the operators are
then interpreted as quantum observables. So, one of the 
main steps of the canonical quantization formalism 
consists basically in replacing the classical observables 
of the physical system by operators. 
Commutation relations are then imposed on the operators 
in order to be in  agreement with Heisenberg's uncertainty 
relation. 
Although this procedure is widely used in quantum 
mechanics and quantum field theory, it is
still far from being completely understood. 
In particular, one expects that in certain limit
the quantum system reduces to the classical one. 
This is usually done by applying the correspondence 
principle, according to which the quantum commutator 
must lead to the classical Poisson bracket when 
Planck's constant vanishes. However, it is known 
that in general this limit is not well-defined and
inconsistent \cite{gro}. 

One of the main successes of deformation quantization 
consists in providing the correct implementation of
the correspondence principle. Indeed, deformation quantization
avoids the use of operators and, instead, concentrates
on the algebra of classical observables 
(see \cite{wein,zac,waldmann} for recent reviews, 
and \cite{hhajp} for an elementary review.)
Instead of the usual (classical) point multiplication between
observables, a star-product is introduced that takes into 
account the non-local character of quantum observables
and reduces to the classical Poisson bracket in the 
appropriate limit. Whereas the classical observables 
build a commutative algebra with respect to the point
product, the same set forms a noncommutative algebra
with respect to the star-product. Thus, it is not necessary
to introduce new entities (operators) instead of the 
classical observables. The quantum observables are
represented by the same functions on phase space as
the classical observables. The first applications 
of deformation quantization concerned non-relativistic
quantum mechanics \cite{bayen}, but this situation has 
changed dramatically in the past few years. This 
formalism has found many uses in perturbative
and nonperturbative quantum field theory 
\cite{dito,kon,dut,bhw,cat,hugo1,hhqft},
quantum gravity \cite{ant},  
as well as in string theory \cite{hugo2,minic}.

In this work we present a first attempt to apply 
the main concepts of deformation quantization to 
the case of linearized gravity. Our approach consists
basically in representing Einstein's linearized 
equations as a field theory of a metric on the 
background of Minkowski spacetime. This approach
allows us to 
apply in this case the procedure of deformation 
quantization developed for free fields. 
Our approach is also appropriate for getting rid
of the concern regarding the quantization of a
quantity (the perturbation metric) which is 
first considered as infinitesimal in order for
Einstein's linearization to be valid. Indeed, 
intuitively one expects that quantum effects become  
important only when the gravitational field is high
enough, and the perturbation metric in the standard
approach does not behave this way. In the field
theoretical approach the metric is arbitrary and
linearized Einstein's equations follow from a 
variational principle. The corresponding Lagrangian
is regular and the Hamiltonian turns out to be
equivalent to the Hamiltonian of an infinite sum 
of harmonic oscillators. 

This paper is structured as follows. In Section
\ref{sec2} we derive Einstein's linearized 
equations in the standard manner and present
an alternative field theoretical approach 
based upon the analogy between electrodynamics
and linearized gravity. In Section \ref{sec3},
we review the main aspects of deformation quantization
and calculate explicitly the set of quantum states
and the energy spectrum of linearized gravity.
Finally, in Section \ref{conclusions} we discuss our results.

\section{Linearized gravity}
\label{sec2}

In this section we first review the standard 
approach of linearized gravity in which the
field equations are obtained by linearizing
Einstein's equations. Then we present an alternative
approach based on  the usual variational procedure of 
field theory and obtain the corresponding
Hamiltonian.

\subsection{Linearized Einstein's equations}
\label{seclee}

In most textbooks on general relativity, linearized gravity is
considered at the level of the field equations. Indeed,
in the usual approach to gravity, one starts from
the Einstein-Hilbert action \cite{wald}
\be
S= \int R \sqrt{-g} d^4 x + \alpha_m \int {\cal L}_m d^4x \ ,
\label{eh}
\ee
where $R$ is the curvature scalar associated to the metric
$g_{\alpha\beta}, \ (\alpha,\beta,... =0,1,2,3)$ of spacetime,
$\alpha_m$ is a coupling constant and ${\cal L}_m$ is 
the Lagrangian density that represents the matter contents in 
spacetime. The variation of (\ref{eh}) with respect to the metric
yields the Einstein equations 
\be 
R_{\alpha\beta} -{1\over 2}  g_{\alpha\beta} R
= 8\pi T_{\alpha\beta} \ , 
\label{ein}
\ee
where the energy-momentum tensor 
is defined  in terms of the variational derivative as
\be
 T_{\alpha\beta}=-{\alpha_m\over 8\pi}{1\over \sqrt{ -g}}
{\delta {\cal L}_m \over   \delta g^{\alpha\beta}} \ .
\label{emt}
\ee
In the weak field approximation
of linearized gravity one imposes the metric $g_{\alpha\beta}=
\eta_{\alpha\beta} + h_{\alpha\beta}$ such that $h_{\alpha\beta}
<< \eta_{\alpha\beta}$ is an infinitesimal perturbation
of the background Minkowski metric, $\eta_{\alpha\beta}$.
In particular, one can choose a Cartesian-like coordinate
system in which $\eta_{\alpha\beta}={\rm diag}(-,+,+,+)$
and $|h_{\alpha\beta}|<<1$. If we now 
consider the first order approximation of the left-hand side 
of (\ref{ein}) with respect to $h_{\alpha\beta}$, and impose
the Lorentz gauge condition 
\be
\overline h ^{\alpha\beta}_{\ \ ,\beta} = 0 \ , 
\qquad {\rm with} \qquad 
\overline h^{\alpha\beta} = h^{\alpha\beta} - 
{1\over 2}\eta^{\alpha\beta} h \  , \qquad 
h=\eta_{\alpha\beta}h^{\alpha\beta}
\label{lor}
\ee
then Einstein's equations reduce to
\be
\overline h _{\alpha\beta,\gamma}^{\ \ \ \ \gamma} 
= - 16 \pi T_{\alpha\beta} \ .
\label{lin}
\ee

\subsection{A field theoretical approach}
\label{secfield}

An alternative approach for deriving the linearized equations
(\ref{lin}), in which the particular aspects of a field theory
become more plausible, consists in using the analogy between
Maxwell's equations of electromagnetism and linearized Einstein's
equations. To show this analogy explicitly let us consider 
a non-zero vector $U^\alpha$ and define 
\be
A_\alpha =- {1\over 4} \overline h_{\alpha\beta}U^\beta\ ,
\quad {\rm and} \quad 
J_\alpha =-T_{\alpha\beta}U^\beta \ .
\label{def}
\ee
Calculating the D'Alembertian $A_{\alpha,\gamma}^{\ \ \ \gamma}$,
one can see immediately that the Maxwell equations 
\be
A_{\alpha,\gamma}^{\ \ \ \gamma} = - 4 \pi J_\alpha \ ,
\label{max}
\ee
are equivalent to the linearized Einstein equations (\ref{lin})
if the components of the arbitrary vector $U^\alpha$ satisfy
 the equations
\be
\overline h _{\alpha\beta}U^{\beta\ \ \gamma}_{\ ,\gamma}
+ 2 \overline h _{\alpha\beta,\gamma} U^{\beta,\gamma} = 0 \ .
\label{con1}
\ee
On the other hand, the Lorentz gauge condition (\ref{lor}) turns
out to be equivalent to the condition $A^\alpha_{\ ,\alpha}=0$ if 
the equation 
\be 
\overline h _{\alpha\beta}U^{\beta,\alpha} = 0 \ ,
\label{con2}
\ee
is satisfied. In this way we see that the linearized 
Einstein equations can be written in the Maxwell-like form
(\ref{max}) (in the Lorentz gauge) 
by introducing the additional vector $U^\alpha$ 
which is required to satisfy the conditions (\ref{con1})
and (\ref{con2}). For a given metric $\overline h _{\alpha\beta}$,
these conditions represent a system
of four second order partial differential  (\ref{con1})
and one first order partial differential equation (\ref{con2})
for the four components $U^\alpha$. So one should guarantee
the existence of solutions to this system for the Maxwell-like
representation (\ref{max}) to be valid. Obviously, 
the trivial vector 
$U^\alpha=$ const satisfies these requirements. In principle,
more general solutions might exist, but in this work we will 
restrict ourselves to this special case since it is sufficient
for our purposes. 

On the other hand, it is well known that Maxwell equations
(\ref{max}) 
can be obtained by varying the Lagrangian (density)
\be
{\cal L}_{Max}= -{1\over 4} F_{\alpha\beta}F^{\alpha\beta}
+ 4 \pi A_\alpha J^\alpha \ , \quad {\rm where}\quad
F_{\alpha\beta}= A_{\beta,\alpha}-A_{\alpha,\beta} \ ,
\label{lmax}
\ee
with respect to the potential $A_\alpha$. Let us now try to
construct the corresponding Hamiltonian formalism. 
The configuration variables 
are given by the set of components $A_\alpha$. 
For the corresponding
canonically conjugate momenta we obtain
$\Pi_\alpha = 
\partial {\cal L}_{Max} / 
\partial \dot A_\alpha = F_0^{\ \alpha}$,
where a dot denotes the derivative with respect
to the time coordinate $x^0$. Then $\Pi_0 = 0$
and consequently we have a singular Lagrangian
from which a Hamiltonian cannot be constructed. 
In field theory the quantization of such Lagrangians
is performed by using Dirac's method for 
systems with constraints (see, for instance, 
\cite{gt}). But in the context of 
deformation quantization this method 
is still under construction \cite{dqconst}. 
In the case of Maxwell's theory, however,
an alternative approach exists \cite{bs} 
that consists
in modifying the original Maxwell Lagrangian according
to (we take $J_\alpha=0$ for simplicity)
\be 
{\cal L} = -{1\over 4} F_{\alpha\beta}F^{\alpha\beta}
-{1\over 2} (A_\alpha^{\ ,\alpha})^2 \ .
\label{lmod0}
\ee
The field equations are again $A_{\alpha,\beta}^{\ \ \ \beta} =0$
and the Lorentz gauge condition $A_\alpha^{\ \ ,\alpha}=0$
has to be postulated separately.
After some algebraic manipulations, the Lagrangian 
(\ref{lmod0}) can be rewritten as
\be
{\cal L} = -{1\over 2} A_{\alpha,\beta}A^{\alpha,\beta} 
+{1\over 2} \Lambda^\beta_{\, ,\beta} \ , \qquad
\Lambda^\beta = A^\beta_{\, ,\gamma} A^\gamma 
- A^\beta A^\gamma_{\, ,\gamma} \ .
\label{lmod}
\ee
The second term can be neglected as it can be transformed after
integration into a surface term that does not contribute to the
field equations.
But the main point about the Lagrangian (\ref{lmod}) is that
it is regular. 
In fact, it can easily be seen that $\Pi_\alpha
= \dot A_\alpha$ and the corresponding Hamiltonian is
\be
{\cal H} = {1\over 2} (\Pi_\alpha\Pi^\alpha+A_{\alpha, i}
A^{\alpha, i} )\ .
\label{ham}
\ee
Then the canonical variables of the phase space satisfy the 
canonical  
commutation relations with respect to the Poisson bracket:
\be
\{A_\alpha, \Pi_\beta\}= \delta_{\alpha\beta} \ , \quad
\{A_\alpha, A_\beta\}=\{\Pi_\alpha, \Pi_\beta\} = 0\ .
\ee
We will use the Hamiltonian (\ref{ham}) in the context
of deformation quantization in section \ref{sec3}.
Finally, let us mention that the Lagrangian (\ref{lmod})
is invariant with respect to the transformation
\be
A_\alpha \rightarrow A'_\alpha = A_\alpha + \Sigma_{,\alpha}\ ,
\quad {\rm with} \quad
\Sigma_{,\alpha}^{\ \ \alpha} = 0 \ .
\label{freedom}
\ee
This is a special gauge transformation that can be used
to eliminate non-physical degrees of freedom. 

The main point of this alternative approach is that
we now can ``forget'' that ${\cal L}$ is an approximate 
Lagrangian and proceed as in standard classical field
theory. That is, (\ref{lmod}) can be interpreted
as the Lagrangian for the metric field $h_{\alpha\beta}$
which is defined on the Minkowski spacetime with metric
$\eta_{\alpha\beta}$. So we are dealing with a standard
field theory in which the background Minkowski metric
does not interact with the field $h_{\alpha\beta}$ that
now can be completely arbitrary, i.e. it is not necessarily 
an infinitesimal perturbation of the background metric.
In this manner we can avoid the concern mentioned in the 
introduction
about the quantization of an infinitesimal quantity.

\section{Deformation quantization}
\label{sec3}

The classical description of the evolution of 
a physical system is usually represented in the 
phase space $\Gamma$, which is a manifold of even 
dimension. If a (non-degenerate) symplectic two-form 
 $\alpha$ exists on $\Gamma$, then the phase space 
is called a symplectic space. 
Observables are real valued functions defined on 
the phase space: $f,g : \Gamma \rightarrow R$. 
With respect to the usual point multiplication
$(fg)(x) = f(x)g(x)$, where $x =(x^1, x^2, ... x^{2n})$
is a set of coordinates on (an open subspace of) $\Gamma$,
the observables build a commutative algebra. 
The symplectic structure $\alpha$ allows us
to introduce the Poisson bracket of observables as
$\{f,g\}(x) = \alpha^{ab}\partial_a f(x) \partial_b g(x)$,
where $\partial_a \ (a = 1,2,... 2n)$ is the 
(covariant) derivative in $\Gamma$. 
With respect to the Poisson bracket 
the set of observables build a Lie-Poisson algebra.
The equations of motion in phase space 
acquire a particular symmetric form in terms of 
the Poisson brackets 
$\dot x ^a= \{ x ^a, {\cal H}\}$, 
a relationship which is valid for any function
of  phase space coordinates.

In the canonical approach to quantization one replaces
the observables by (self-adjoint) 
operators which act on the Hilbert space. The physical 
states are vectors in the Hilbert space. Poisson brackets
are replaced by commutators which, when applied to 
the operators associated with the basic observables 
in phase space, satisfy the canonical commutation 
relations. 
Despite its great success especially in the perturbative
approach  to the physics of elementary particles, 
this procedure is still not completely understood. 
The passage from functions to operators is one 
important step in the canonical approach and despite
many efforts done to explain it, today the best way
to avoid all kind of existence proofs and mathematical 
difficulties is just to 
assume it as a postulate. 

Deformation quantization is 
essentially an attempt to avoid the passage from 
functions to operators. In fact, it focuses on 
the algebra of observables of the phase space
and replaces the usual point product of functions
by a star-product. The canonical
commutation relations are now a consequence of 
the definition of the star-product. 
An important advantage of this procedure is that 
quantum as well classical observables are functions
defined on the phase space and no operators are
required.
From the mathematical point of view, 
the deformation quantization of
a given classical system consists in giving 
an appropriate definition of the star-product 
which acts (on functions defined) on the phase 
space. In physics, however, to understand
a quantum system one needs to know its quantum 
states and their energy spectrum. To this end,
deformation quantization postulates the 
existence of a time-evolution function, Exp($Ht$),
which satisfies the differential equation \cite{hhqft}
\be
i \hbar {d\over dt} {\rm Exp}(Ht) = H * {\rm Exp}(Ht)
 \ , \label{schroe}
\ee
where $H$ is the Hamiltonian of the classical system.
Moreover, it is assumed that the time-evolution function
allows a Fourier-Dirichlet expansion as
\be
{\rm Exp}(Ht) = \sum_E \pi_E e^{-itE/\hbar} ,
\label{exp}
\ee
where $E$ is the energy (a real number) 
associated with the state $\pi_E$ 
(distribution on the phase space), or Wigner function, which
satisfies the so called *-genvalue equation
\be
H*\pi_E = E \pi_E \ .
\label{eigen}
\ee
The states are idempotent and complete. i.e.:
\be
\pi_E*\pi_{E'} = \delta_{E,E'} \pi_E \ , \qquad
\sum_E \pi_E = 1 \ .
\ee

As a consequence, the spectral decomposition of the Hamiltonian
is give as
\be
 H = \sum_E \, E \, \pi_E \ .
\ee
Essentially,  the objects that are necessary for carrying out
the deformation quantization of a physical system 
are the {\it classical} Hamiltonian $H$ and
the *-product. For a given phase space it is not 
clear {\it a priori} if a consistent *-product exists or not
and, for a general phase space, this is still an open problem
\cite{waldmann}. 
In the case of free (non-interacting) fields that 
can be considered heuristically as the sum of an infinite 
number of harmonic oscillators, it has been shown \cite{dito} 
that the normal star-product is the only admissible 
star-product. The normal *-product between two functions
$f$ and $g$ on phase space is defined by 
\be
f*_N g = e^{N_{12}} f(a^{(1)},\overline a ^{(1)})
g(a^{(2)},\overline a ^{(2)})\bigg|_{a^{(1)}=a^{(2)}=a} \ ,
\qquad
N_{12} = \hbar \delta_{ij}{\partial\over \partial a_i^{(1)} }
{\partial\over \partial \overline a_j^{(2)} } \ ,
\label{norpro}
\ee
where the superscritpts (1) and (2) denote two arbitrary points
in phase space and $a = (a_1, a_2, ..., a_n)$. Furthermore, an 
overline denotes complex conjugation. The set of phase space
variables has to satisfy the standard commutation relations 
with respect to the Poisson bracket, i.e. 
$\{a_i,\overline a _j\} = \delta_{ij}, \ \{a_i, a _j\}=
\{\overline a_i,\overline a _j\}=0.$ In particular 
one can choose  $a_j=1/\sqrt{2} (
x_j+ix_{n+j}), (j=1,...,n)$, 
where we are assuming that the configurational variable 
$x_j$ and its conjugate 
momentum $x_{n+j}$ have been normalized to come out with the same 
units. 

Let us now apply this procedure to the linearized theory.
According to the results given in section \ref{sec2}, the canonical
variables in the phase space of linearized gravity 
are the potentials $A_\alpha=-(1/4)\overline h _{\alpha\beta} U^\beta$ 
and their canonical momenta $\Pi_\alpha =\dot A_\alpha$. 
The Hamiltonian is given by
\be
H ={1\over 2}\int d^3x \left(\Pi_\alpha\Pi^\alpha 
+A_{\alpha,i}A^{\alpha,i}\right) \ .
\label{hamc}
\ee
The main step in the quantization procedure consists 
in solving Eq.(\ref{schroe}) by using the normal *-product
(\ref{norpro}).  To this end it is convenient to change
from the variables of phase space $(A_\alpha, \Pi_\alpha)$
to a new set of canonically conjugate variables in which 
the Hamiltonian takes the simplest possible form. This
procedure is very well known in field theory and consists
in introducing the momentum representation of the
phase variable $A_\alpha$ according to \cite{bs}
\be
A_\alpha(x)= {1\over (2\pi)^{3/2}}\int {d{\bf k}\over \sqrt{2k_0}}
\left[a_\alpha({\bf k})e^{ikx} + 
\overline a_\alpha({\bf k})e^{-ikx}\right]
\ , 
\label{momentum}
\ee
where $k$ is a null vector $k_\mu k^\mu = -k_0^2 +{\bf k}^2 =0, $
and $kx=k_\mu x^\mu$. Then
\be
\Pi_\alpha=\dot A_\alpha =
{i\over (2\pi)^{3/2}}\int d{\bf k} \sqrt{k_0\over 2}
\left[a_\alpha({\bf k})e^{ikx} -
\overline a_\alpha({\bf k})e^{-ikx}\right] \ ,
\label{pis}
\ee
and
\be 
A_{\alpha, j} = 
{i\over (2\pi)^{3/2}}\int {d{\bf k}\over \sqrt{2k_0}} k_j
\left[a_\alpha({\bf k})e^{ikx} -
\overline a_\alpha({\bf k})e^{-ikx}\right] \ .
\label{aaa}
\ee
From the commutation relations for $A_\alpha$ and $\Pi_\beta$ 
it can be shown that
\be
\{a_\alpha, \overline a_\beta\} = \delta_{\alpha\beta}\ ,
\quad 
\{a_\alpha, a_\beta\} =
\{\overline a_\alpha, \overline a_\beta\} = 0\ ,
\ee
where the same value of ${\bf k}$ has been assumed in all the 
arguments. Introducing  (\ref{pis}) and (\ref{aaa}) into the
Hamiltonian (\ref{hamc}) and performing some of the integrations
we obtain
\be
H={1\over 2} \int k_0 \eta^{\alpha\beta}
a_\alpha({\bf k})\overline a_\beta({\bf k}) d{\bf k} 
\label{hamas}
\ee
We see that the resulting Hamiltonian is linear 
in the new variables and does not contain derivatives.
It can be interpreted as an infinite sum of harmonic
oscillators. This is an important observation 
\cite{dito,hugo1,hugo2}
that allows us to formally apply the normal *-product
as defined in (\ref{norpro}). 
In fact, when going
from a system with a finite number of degrees of freedom
to a field theory, one only has to ``replace'' 
partial derivatives by variational derivatives. 
We use this fact to calculate time-evolution function  
as the solution of Eq.(\ref{schroe}). Then we have
\be
\ln {\rm Exp}_N (Ht) = {1\over \hbar}\left( e^{-ik_0t} -1\right)
\int \eta^{\alpha\beta} a_\alpha({\bf k})\overline a_\beta({\bf k}) d{\bf k} \ ,
\label{tef}
\ee
where the subscript $N$ indicates that in Eq.(\ref{schroe}) the normal 
*-product
has been used. Using the definition of the exponential of a functional,
the last expression can be written as
\be
{\rm Exp}_N (Ht) = \exp\left( -{1\over \hbar}\int \eta^{\alpha\beta}
a_\alpha({\bf k})\overline a_\beta({\bf k}) d{\bf k}\right)
\sum_{n=0}^\infty { e^{-ink_0 t} \over n!\hbar^n}\int \eta^{\alpha\beta}
a^n_\alpha({\bf k})\overline a^n_\beta({\bf k}) d{\bf k}
\ .
\label{tef1}
\ee
Comparing this expression with the Fourier-Dirichlet expansion 
(\ref{exp}) we can identify the corresponding states as
\be
\pi^N_{E_0} = \exp\left(-{1\over \hbar} \int  \eta^{\alpha\beta} 
a_\alpha({\bf k})\overline a_\beta({\bf k}) d{\bf k}\right) 
\ ,
\label{state0}
\ee
\be
\pi^N_{E_n} = {1\over n! \hbar^n} \pi^N_{E_0} 
\int  \eta^{\alpha\beta} 
a^n_\alpha({\bf k})\overline a^n_\beta({\bf k}) d{\bf k}
\ ,
\label{staten}
\ee
and the energy spectrum
\be
E_n = n\hbar k_0 \ .
\label{spectrum}
\ee
In this manner we have arrived at the main result of
quantization: The determination of the quantum states
and the energy spectrum of the system. The main advantage
of deformation quantization consists in achieving this 
goal without using the operator formalism. 
Now we are confronted with the problem of finding the
physical significance of our results. 
To this end, let us remember that at the classical 
level we have derived Einstein's linearized equations
directly from the Lagrangian (\ref{lmod}), and have noticed
that the Lorentz gauge conditions have to be postulated 
as an additional requirement. It seems therefore natural
to impose this requirement on the quantum states 
to find out which of them are physical. 
From Eq.(\ref{momentum}) we find that the Lorentz gauge
conditions are equivalent to
\be
A_\alpha^{\ ,\alpha} = 
 {1\over (2\pi)^{3/2}}\int {d{\bf k}\over \sqrt{2k_0}} i k^\alpha
\left[a_\alpha({\bf k})e^{ikx} - 
\overline a_\alpha({\bf k})e^{-ikx}\right] = 0 \ .
\label{qlor}
\ee
Clearly, this condition is identically satisfied if 
\be
k^\alpha a_\alpha({\bf k}) = 0 \ ,
\label{ort}
\ee
what implies that only three components of $a_\alpha({\bf k})$
are linearly independent. Furthermore, the gauge freedom 
given by Eq.(\ref{freedom}) implies that the harmonic 
function $\Sigma(x)$ can be used to eliminate an
additional component of $a_\alpha({\bf k})$. So we are 
left with only two true components, say, 
$a_1({\bf k})$ and $a_2({\bf k})$. This is in accordance
with the fact that gravitational fields possess only two
physical degrees of freedom \cite{wald}. 
From Eqs.(\ref{state0}) and (\ref{staten}) we see that all
states are specified as powers of $a_1({\bf k})$ and $a_2({\bf k})$.
The results should not depend on the choice of these
two linear independent components, but one can 
use these freedom to adapt the formalism to different
physical situations. For instance, in the case 
of the Newtonian limit 
it seems reasonable to choose the Newtonian potential
$\phi$ and one of the ``gravitomagnetic'' functions \cite{mtw},
say $\gamma$, as independent configuration variables
so that $A_\alpha = -(1/4)(4\phi, \gamma, 0, 0)$. 
In the case of gravitational waves a more suitable
choice would be $A_\alpha= -(1/4)(0, \gamma_1, \gamma_2, 0)$
where $\gamma_1$ and $\gamma_2$ are now related to the 
special combination of the metric components that describe
gravitational waves (see, for instance, \cite{mtw}).

Independently of the choice of gauge, the states and 
spectrum are represented by Eqs.(\ref{state0}),
(\ref{staten}) and (\ref{spectrum}). The coefficients
$a_\alpha({\bf k})$ can be interpreted as densities
that determine distributions in phase space, i.e.
as state densities. But the formalism of deformation 
quantization allows a transition to the operator
formalism according to certain fixed rules \cite{hhqft}.
In that case, one would expect that the operator
counterparts of $a_\alpha$ and $\overline a_\alpha$ 
would correspond to the annihilation and creation
operators of standard quantum field theory.
The energy spectrum (\ref{spectrum}) is discrete 
with vanishing zero-point energy. If we would use the 
Moyal product for the quantization, we would obtain
a non-vanishing zero point energy and would be 
confronted with the problem of divergencies that
commonly appears in perturbative quantum field
theory \cite{dito}.

\section{Conclusions}
\label{conclusions}

The aim of this work was to apply the formalism of deformation
quantization to linearized Einstein's equations. 
We first show two alternative ways to consider Einstein's
linearized equations in a field theoretical 
approach. We then use the modified Maxwell representation of 
linearized gravity to calculate the classical Hamiltonian 
of the theory, avoiding the problem of a singular Lagrangian. 
The Hamiltonian is one of the main ingredients necessary to carry out 
the deformation quantization of any physical system. 
We use the normal star-product to derive the commutation
relations, in analogy with other free (linear) fields.
The expression for the time-evolution function is found 
explicitly, and the Fourier-Dirichlet expansion of the Hamiltonian
is used to derive the energy spectrum and the complete
set of states of the system. A more detailed analysis
is necessary in order to clarify further 
 the physical 
meaning of the states. We have used the momentum 
representation in analogy with the standard methods
of quantum field theory. For this reason, the 
results of the quantization are more adapted to 
a possible interpretation in terms of elementary particles and
not in terms of a possible quantization of space and time.
This, however, is a much more complicated problem
that requires a separate and detailed  treatment.

% related to conceptual issues of quantum gravity, 
% a theory that still has to be formulated. 

\section*{Acknowledgments}

It is a great pleasure to dedicate this work to Alberto 
 Garc\'\i a on his 60-th birthday. 
 This work was in part supported by
DGAPA-UNAM grant IN112401, CONACyT grant
36581-E, and US DOE grant DE-FG03-91ER40674.
H.Q. thanks UC-MEXUS for
support.


\begin{thebibliography}{99}

\bibitem{gro} H. J. Groenewold, Physica (Amsterdam) {\bf 12}, 405 (1946);
L. van Hove, Proc. R. Acad. Sci. Belgium {\bf 26}, 1 (1951). 

\bibitem{wein} A. Weinstein, Semin. Bourbaki, Asterique {\bf 789}, 389 (1995).

\bibitem{zac} C. Zachos, hep-th/0110114. 

\bibitem{waldmann} S. Waldmann, hep-th/0303080.

\bibitem{hhajp} A. C. Hirshfeld and P. Henselder, Am. J. Phys. {\bf 70}, 
537 (2002).

\bibitem{bayen} F. Bayen, M. Flato, C. Fronsdal, A. Lichnerowicz, 
and D. Sternheimer, Ann. Phys. (N.Y.) {\bf 111}, 61 (1978).

\bibitem{dito} J. Dito, Lett. Math. Phys. {\bf 20}, 125 (1990).

\bibitem{kon} M. Kontsevich, q-alg/9709040. 

\bibitem{dut} M. D\"utch and K. Fredenhagen, hep-th/9807215, 
hep-th/0101079.

\bibitem{bhw} M. Bordemann, H.C. Herbig and S. Waldmann, 
Comm. Math. Phys. {\bf 210}, 107 (2000).

\bibitem{cat} A. S. Cattaneo and G. Felder, 
Comm. Math. Phys. {\bf 212}, 591 (2000).


\bibitem{hugo1} H. Garcia-Compean, J.F. Plebanski, M. Przanowski
and F.J. Turrubiates, Int. J. Mod. Phys. A {\bf 16}, 2533 (2001).


\bibitem{hhqft} A. C. Hirshfeld and P. Henselder, Ann. Phys. {\bf 298},
382 (2002).

\bibitem{ant} F. Antonsen, gr-qc/9712012.



\bibitem{hugo2} H. Garcia-Compean, J.F. Plebanski, M. Przanowski
and F.J. Turrubiates, J. Phys. A {\bf 33}, 7935 (2000).

\bibitem{minic} D. Minic, hep-th/9909022.


\bibitem{dqconst} F. Antonsen, gr-qc/9710021.

\bibitem{wald} R. M. Wald, {\it General Relativity} 
(The University of Chicago Press, Chicago, 1984).


\bibitem{mtw} C. W. Misner, K. S. Thorne, and J. A. Wheeler, 
{\it Gravitation} (Freeman Ed., San Francisco, 1973).


\bibitem{gt} D. M. Gitman and I. V. Tyutin, {\it Quantization of 
fields with constraints} (Springer-Verlag, Berlin, 1990).
 

\bibitem{bs} N. N. Bogoliubov and D. V. Shirkov, {\it Quantum fields}
(Benjamin Ed., USA, 1983)








\end{thebibliography}
\end{document}